\documentstyle[12pt,fleqn]{article}
\def\beq{\begin{equation}}
\def\eeq{\end{equation}}

\begin{document}

\vspace*{15mm}
\begin{center}
{\Large Separability Criterion for Density Matrices} \\[2cm]
Asher Peres$^*$ \\[7mm]
{\sl Department of Physics, Technion---Israel Institute of
Technology, 32\,000 Haifa, Israel}
\end{center}\vfill

\noindent{\bf Abstract}\bigskip

A quantum system consisting of two subsystems is {\it separable\/} if
its density matrix can be written as $\rho=\sum_A w_A\,\rho_A'\otimes
\rho_A''$, where $\rho_A'$ and $\rho_A''$ are density matrices for the
two subsytems. In this Letter, it is shown that a necessary condition
for separability is that a matrix, obtained by partial transposition of
$\rho$, has only non-negative eigenvalues. This criterion is stronger
than Bell's inequality. \vfill

\vfill

\noindent $^*$\,Electronic address: peres@photon.technion.ac.il

\vfill\newpage

A striking quantum phenomenon is the inseparability of composite quantum
systems. Its most famous example is the violation of Bell's inequality,
which may be detected if two distant observers, who independently {\it
measure\/} subsytems of a composite quantum system, {\it report\/} their
results to a common site where that information is analyzed~[1].
However, even if Bell's inequality is satisfied by a given composite
quantum system, there is no guarantee that its state can be {\it
prepared\/} by two distant observers who receive {\it instructions\/}
from a common source. For this to be possible, the density matrix $\rho$
has to be separable into a sum of direct products,

\beq \rho=\sum_A w_A\,\rho_A'\otimes\rho_A'', \label{sep}\eeq
where the positive weights $w_A$ satisfy $\sum w_A=1$, and where
$\rho_A'$ and $\rho_A''$ are density matrices for the two subsystems. A
separable system always satisfies Bell's inequality, but the converse is
not necessarily true~[2--5]. In this Letter, I shall derive a simple
algebraic test, which is a {\it necessary\/} condition for the existence
of the decomposition (\ref{sep}). This criterion is more restrictive
than Bell's inequality, or than the $\alpha$-entropy inequality~[6].

The derivation of this separability condition is best done by writing
the density matrix elements explicitly, with all their indices~[1]. For
example, Eq.~(\ref{sep}) becomes

\beq \rho_{m\mu,n\nu}=
  \sum_A w_A\,(\rho'_A)_{mn}\,(\rho''_A)_{\mu\nu}. \eeq
Latin indices refer to the first subsystem, Greek indices to the second
one (the sub\-systems may have different dimensions). Note that this
equation can always be satisfied if we replace the quantum density
matrices by classical Liouville functions (and the discrete indices are
replaced by canonical variables, {\bf p} and {\bf q}). The reason is
that the only constraint that a Liouville function has to satisfy is
being non-negative. On the other hand, we want quantum density matrices
to have non-negative {\it eigenvalues\/}, rather than non-negative
elements, and the latter condition is more difficult to satisfy.

Let us now define a new matrix,

\beq \sigma_{m\mu,n\nu}\equiv\rho_{n\mu,m\nu}. \eeq
The Latin indices of $\rho$ have been transposed, but not the Greek
ones. This is not a unitary transformation but, nevertheless, the
$\sigma$ matrix is Hermitian. Moreover, its eigen\-values are invariant
under separate unitary transformations of the bases used by the two
observers. Indeed if

\beq \rho\to (U'\otimes U'')\,\rho\,(U'\otimes U'')^\dagger, \eeq
we have

\beq \sigma\to
 (U'^T\otimes U'')\,\sigma\,(U'^T\otimes U'')^\dagger, \eeq
which also is a unitary transformation, leaving the eigenvalues of
$\sigma$ invariant.

When Eq.~(\ref{sep}) is valid, we have

\beq \sigma=\sum_A w_A\,(\rho_A')^T\otimes\rho_A''. \label{sig}\eeq
Since the transposed matrices $(\rho'_A)^T\equiv(\rho'_A)^*$ are
non-negative matrices with unit trace, they can also be legitimate
density matrices. It follows that {\it none of the eigenvalues of
$\sigma$ is negative\/}. This is a necessary condition for
Eq.~(\ref{sep}) to hold.

As an example, consider a pair of spin-$1\over2$ particles in a Werner
state (an impure singlet), consisting of a singlet fraction $x$ and
a random fraction $(1-x)$~[7]. Note that the ``random fraction'' $(1-x)$
also includes singlets, mixed in equal proportions with the three
triplet components. We have

\beq \rho_{m\mu,n\nu}=x\,S_{m\mu,n\nu}+
  (1-x)\,\delta_{mn}\,\delta_{\mu\nu}\,/4,\eeq
where the density matrix for a pure singlet is given by

\beq S_{01,01}=S_{10,10}=-S_{01,10}=-S_{10,01}=\mbox{$1\over2$}, \eeq
and all the other components of $S$ vanish. (The indices 0 and 1 refer
to any two ortho\-gonal states, such as ``up'' and ``down.'') A
straightforward calculation shows that $\sigma$ has three eigenvalues
equal to $(1+x)/4$, and the fourth eigenvalue is $(1-3x)/4$. This lowest
eigenvalue is positive if $x<{1\over3}$, and the separability criterion
is then fulfilled. This result may be compared with other criteria:
Bell's inequality holds for $x<1/\sqrt{2}$, and the $\alpha$-entropic
inequality~[6] for $x<1/\sqrt{3}$. These are therefore much weaker tests
for detecting inseparability than the condition that was derived here.

In this particular case, it happens that this necessary condition is
also a sufficient one. It is indeed known that if $x<{1\over3}$ it is
possible to write $\rho$ as a mixture of unentangled product states~[8].
This result suggests that the necessary condition derived above
($\sigma$ has no negative eigenvalue) might also be sufficient for any
$\rho$. Some time after this Letter was submitted for publication, a
proof of this conjecture was indeed obtained~[9] for composite systems
having dimensions $2\times2$ and $2\times3$. However, for higher
dimensions, the present necessary condition was shown {\it not\/} to be
a sufficient one.

As a second example, consider a mixed state introduced by Gisin~[5].
With the present notations, it consists of a fraction $x$ of the pure
state $a|01\rangle+b|10\rangle$ (with $|a|^2+|b|^2=1$), and fractions
$(1-x)/2$ of the pure states $|00\rangle$ and $|11\rangle$. The
non-vanishing elements of $\rho$ thus are

\beq \rho_{00,00}=\rho_{11,11}=(1-x)/2,\eeq
\beq \rho_{01,01}=x|a|^2, \eeq
\beq \rho_{10,10}=x|b|^2, \eeq
\beq \rho_{01,10}=\rho_{10,01}^*=xab^*. \eeq
It is easily seen that the $\sigma$ matrix has a negative determinant,
and therefore a negative eigenvalue, when 

\beq x>(1+2|ab|)^{-1}. \eeq
This is a lower limit than the one for violation of Bell's inequality,
which requires~[5]

\beq x>[1+2|ab|(\sqrt{2}-1)]^{-1}. \eeq

An even more striking example is the mixture of a singlet and a
maximally polarized pair:

\beq \rho_{m\mu,n\nu}=x\,S_{m\mu,n\nu}+
  (1-x)\,\delta_{m0}\,\delta_{n0}\,\delta_{\mu0}\,\delta_{\nu0}.\eeq
For any positive $x$, however small, this state is inseparable, because
$\sigma$ has a negative eigenvalue. On the other hand, the Horodecki
criterion~[10] gives a very generous domain to the validity of Bell's
inequality: $x\leq 0.8$.

The weakness of Bell's inequality is attributable to the fact that the
only use made of the density matrix $\rho$ is for computing the
probabilities of the various outcomes of tests that may be performed on
the subsystems of a {\it single\/} composite system. On the other hand,
an experimental verification of that inequality necessitates the use of
{\it many\/} composite systems, all prepared in the same way.  However,
if many such systems are actually available, we may also test them
collectively, for example two by two, or three by three, etc., rather
than one by one. If we do that, we must use, instead of $\rho$ (the
density matrix of a single system), a {\it new\/} density matrix, which
is $\rho\otimes\rho$, or $\rho\otimes\rho\otimes\rho$, in a higher
dimensional space. It then turns out that there are some density
matrices $\rho$ that satisfy Bell's inequality, but for which
$\rho\otimes\rho$, or $\rho\otimes\rho\otimes\rho$, etc., violate that
inequality~[11].

This result raises a new question: can we get stronger inseparability
criteria by considering $\rho\otimes\rho$, or higher tensor products? It
is easily seen that no further progress can be achieved in this way. If
$\rho$ is separable as in Eq.~(\ref{sep}), so is $\rho\otimes\rho$.
Moreover, the partly transposed matrix corresponding to
$\rho\otimes\rho$ simply is $\sigma\otimes\sigma$, so that if no
eigenvalue of $\sigma$ is negative, then $\sigma\otimes\sigma$ too has
no negative eigenvalue.\bigskip

I am grateful to R. Horodecki and R. Jozsa for pointing out that
Eq.~(\ref{sig}) could be used instead of a longer derivation that
appeared in an earlier version of this Letter. This work was supported
by the Gerard Swope Fund and the Fund for Encouragement of Research.
\clearpage

\frenchspacing \begin{enumerate}
\item A. Peres, {\it Quantum Theory: Concepts and Methods\/}
(Kluwer, Dordrecht, 1993) Chapters 5 and 6.
\item R. F. Werner, Phys. Rev. A {\bf 40}, 4277 (1989).
\item S. Popescu, Phys. Rev. Lett. {\bf 72}, 797 (1994); {\bf 74}, 2619
(1995).
\item N. D. Mermin, in {\it Quantum Mechanics without Observer\/},
edited by R. K. Clifton (Kluwer, Dordrecht, 1996) pp.~57--71.
\item N. Gisin, Phys. Lett. A {\bf 210}, 151 (1996).
\item R. Horodecki, P. Horodecki, and M. Horodecki, Phys. Lett. A {\bf
210}, 377 (1996).
\item Werner considered only the case $x={1\over2}$. These more
general states were introduced by J. Blank and P. Exner, Acta Univ.
Carolinae, Math. Phys. {\bf 18}, 3 (1977).
\item C. H. Bennett, G. Brassard, S. Popescu, B. Schumacher, J.
Smolin, and W. K. Wootters, Phys. Rev. Lett. 76 (1996) 722.
\item M. Horodecki, P. Horodecki, and R. Horodecki, ``On the necessary
and sufficient conditions for separability of mixed quantum states''
(e-print archive: quant-ph/9605038).
\item R. Horodecki, P. Horodecki, and M. Horodecki, Phys. Lett. A {\bf
200}, 340 (1996).
\item A. Peres, ``Collective tests for quantum nonlocality'' (submitted
to Phys. Rev. A, e-print archive: quant-ph/9603023).

\end{enumerate}
\end{document}